\documentclass[11pt,twoside,final]{aipproc}
\layoutstyle{6x9}
\usepackage{graphicx}

\begin{document}

\title{Latest Results From The DODO Survey: Imaging Planets Around White 
Dwarfs}


\author{Emma Hogan}{
  address={Gemini Observatory, Casilla 603, La Serena, Chile}
}

\author{Matt R. Burleigh}{
  address={Department of Physics and Astronomy, University of Leicester, 
           University Road, Leicester, LE1 7RH, UK}
}

\author{Fraser J. Clarke}{
  address={Department of Astrophysics, The Denys Wilkinson Building, 
           University of Oxford, Keble Road, Oxford, OX1 3RH, UK}
}

\begin{abstract}
The aim of the Degenerate Objects around Degenerate Objects (DODO) survey is 
to search for very low mass brown dwarfs and extrasolar planets in wide orbits 
around white dwarfs via direct imaging. The direct detection of such 
companions would allow the spectroscopic investigation of objects with 
temperatures lower ($<500$~K) than the coolest brown dwarfs currently 
observed. The discovery of planets around white dwarfs would prove that such 
objects can survive the final stages of stellar evolution and place 
constraints on the frequency of planetary systems around their progenitors 
(with masses between $1.5-8\,M_{\odot}$, i.e., early B to mid-F). An 
increasing number of planetary mass companions have been directly imaged in 
wide orbits around young main sequence stars. For example, the planets around 
HR~$8799$ and $1$RXS~J$160929.1-210524$ are in wide orbits of $24-68$~AU and 
$330$~AU, respectively. The DODO survey has the ability to directly image 
planets in post-main sequence analogues of these systems. These proceedings 
present the latest results of our multi-epoch $J$ band common proper motion 
survey of nearby white dwarfs.
\end{abstract}

\maketitle

\section{Introduction}

Directly imaging the extrasolar planets found in orbit around solar type stars 
is difficult as these faint companions are too close to their bright parent 
stars. However, an increasing number of planetary mass companions have been 
directly imaged in wide orbits around young main sequence stars. For example, 
three directly imaged companions, with likely masses between 
$5-13\,M_{\rm{Jup}}$ and projected physical separations of $\sim24$, $38$ and 
$68$~AU, were found around the A-type star HR$8799$ \citep{mmb2008}. Another, 
more extreme example is the $\sim8\,M_{\rm{Jup}}$ companion imaged at a 
surprisingly large separation of $\sim330$~AU around a member of the Upper 
Scorpius association, $1$RXS~J$160929.1-210524$ \citep{ljv2008}. All the 
imaged planetary mass companions found to date have been confirmed to be 
common proper motion companions to their parent stars. However, coronagraphy 
and adaptive optics were needed to detect these faint companions. Another, 
perhaps simpler, solution to the problems of contrast and resolution is to 
instead target intrinsically faint stars.

\section{Planets Around White Dwarfs}

White dwarfs are intrinsically faint stars and can be up to $10,000$ times 
less luminous than their main sequence progenitors, significantly enhancing 
the contrast between any companion and the white dwarf. In addition, any 
companion that avoids direct contact with the red giant envelope as the main 
sequence progenitor evolves into a white dwarf, i.e., planets with an initial 
orbital radius larger than $\sim5$~AU, will migrate outwards as mass is lost 
from the central star by a maximum factor of $M_{\rm{MS}}/M_{\rm{WD}}$ 
\citep{j1924}. This increases the projected physical separation between the 
companion and the white dwarf, substantially increasing the probability of 
obtaining a ground based direct image of a planetary mass companion. The 
evolution of planetary systems during the post-main sequence phase is 
discussed in more detail by \citet{dl1998}, \citet*{bch2002}, \citet{ds2002} 
and \citet{vl2007}.

The direct detection of a planetary mass companion in orbit around a white 
dwarf would allow the spectroscopic investigation of very low mass objects 
cooler ($<500$~K) and older ($>$~few Gyr) than previously found. Such a 
discovery would help provide constraints on models for the evolution of 
planets and planetary systems during the final stages of stellar evolution. In 
addition, the age of any substellar and planetary mass companions discovered 
in such a system can be estimated using the white dwarf cooling age and the 
mass and the lifetime of the main sequence progenitor, providing model-free 
benchmark estimates of their mass and luminosity, which could be used to test 
evolutionary models \citep{pjl2006}. Furthermore, as the $1.5-8\,M_{\odot}$ 
progenitor stars of white dwarfs have spectral types of early B, A and mid-F, 
searching for planetary mass companions in orbit around white dwarfs allows 
the examination of a currently inadequately explored region of parameter 
space, supplying new information on the frequency and mass distribution of 
extrasolar planets around intermediate mass main sequence stars. Finally, 
given that white dwarfs evolve from $1.5-8\,M_{\odot}$ progenitor stars, it is 
possible that they harbour more massive planetary companions (cf. the massive 
planets in wide orbits around the A-type star HR~$8799$), increasing the 
chances of directly detecting planets around white dwarfs.

An initial sample of $\sim40$ targets, with total ages (main sequence 
progenitor lifetime plus the white dwarf cooling age) $<4$~Gyr were selected 
from the catalogue of white dwarfs within $20$~pc \citep{hos2002}. One hour 
multi-epoch observations of these white dwarfs were acquired in the $J$ band 
using Gemini North and \textit{NIRI} for equatorial and Northern hemisphere 
targets, and ESO-VLT and \textit{ISAAC} for Southern hemisphere targets, while 
a small number of observations of equatorial targets were acquired using 
Gemini South and \textit{FLAMINGOS}. These one hour $J$ band images have an 
average sensitivity of $J\sim22.5$~mag and a typical image quality of 
$\sim0.6^{\prime\prime}$, without the use of adaptive optics. Due to the large 
number of faint objects in these deep, wide field ($120^{\prime\prime}$) 
images, all targets are observed again after 1--2 years to determine whether 
there are any common proper motion companions to the white dwarf.

The effective temperature, $T_{\rm{eff}}$, the log~$g$ and the mass, 
$M_{\rm{WD}}$, of the white dwarf are taken from the literature (e.g., 
\citealt{blr2001}, \citealt{dbf2005}). The cooling age of a white dwarf, 
$t_{\rm{WD}}$, can be calculated using evolutionary models. When the cooling 
age was unavailable in the literature, models from \citet*{fbb2001}, which use 
$T_{\rm{eff}}$ and log~$g$ values to calculate the cooling age, were used to 
estimate this value. The initial-final mass relation (IFMR) determined by 
\citet{dnb2006}, based on the measurements of a small number of white dwarfs 
found in young open clusters, was used to determine the mass of the main 
sequence progenitor, $M_{\rm{MS}}$, from $M_{\rm{WD}}$. This linear IFMR is 
given as 
\begin{equation} 
M_{\rm{WD}}=0.133\,M_{\rm{MS}}+0.289 
\label{ifmr} 
\end{equation} 
Recent observations of white dwarfs in older open clusters have placed 
constraints on the low mass end of the IFMR, suggesting that this equation is 
valid down to white dwarf masses of $0.54\,M_{\odot}$ \citep{khk2008}. The 
main sequence progenitor lifetime, $t_{\rm{MS}}$, is estimated using the 
equation 
\begin{equation}
t_{\rm{MS}}=10\,\left(\frac{M_{\rm{MS}}}{M_{\odot}}\right)^{-2.5}
\label{mslifetime}
\end{equation}
where $t_{\rm{MS}}$ is measured in Gyr \citep{w1992}. Finally, the 
completeness limit for each image was estimated by determining the magnitude 
at which $90\%$ and $50\%$ of inserted artificial stars were recovered from 
each image. The ``COND'' evolutionary models for cool brown dwarfs and 
extrasolar planets \citep{bcb2003}, along with the magnitudes at which $90\%$ 
and $50\%$ of artificial stars were recovered, were then used to estimate the 
minimum mass and corresponding effective temperature of a companion that could 
be detected in both epoch images.

The total age of the white dwarf is equal to the sum of the main sequence 
progenitor lifetime and white dwarf cooling age, both of which depend upon 
evolutionary models. While the cooling age errors are small and well 
constrained \citep{fbb2001}, and the scatter in the empirical IFMR is 
significantly reducing as more and higher quality observations are made of 
white dwarfs in open clusters \citep{cdn2008}, the main sequence progenitor 
lifetimes rely on models that are difficult to calibrate (e.g., 
\citealt{cig2008}). Therefore, to take these uncertainties into account, a 
conservative error of $\pm25\%$ is applied to the total age of each white 
dwarf (note that the white dwarf cooling age is the dominant timescale for 
most of the targets in the DODO survey). However, at ages $>1$~Gyr, the 
``COND'' evolutionary models indicate that the absolute magnitudes of 
substellar objects are relatively insensitive to changes in their age, 
implying that even with a $\pm25\%$ error, the resulting error on the mass of 
a companion is small (Table~\ref{results}).

\section{Results}

In \citet{hbc2009}, we presented the results of 23 nearby equatorial and 
Northern hemisphere white dwarfs. We ruled out the presence of any common 
proper motion companions, with limiting masses determined from the 
completeness limit of each observation, to 18 white dwarfs. For the remaining 
five targets, the motion of the white dwarf was not sufficiently separated 
from the non-moving background objects in each field. Third epoch images have 
now been obtained for four of these five white dwarfs. These more recent 
observations rule out the presence of any common proper motion companions to 
the four white dwarfs. Since then, five Southern hemisphere white dwarfs have 
been fully analysed and also show no evidence of any common proper motion 
companions \citep{hcb2010}. Including the non-detection of a companion around 
WD~$0046+051$ \citep{bch2008}, a total of 29 white dwarfs from the DODO survey 
have been fully analysed (Table~\ref{results}).
\begin{table}
\caption{Results for 29 white dwarfs from the DODO survey\label{results}}
\begin{tabular}{ccccr@{$^{+}_{-}$}lcr@{ - }lr@{ - }lc}
\hline
\multicolumn{1}{c}{\bf White\tablenote{Columns: $t_{\rm{tot}}$ is the ``COND'' evolutionary model age used; $50\%$ gives the $50\%$ completeness limits in terms of apparent $J$ magnitude, mass, $M$, measured in Jupiter masses, and effective temperature, $T$, measured in Kelvin, respectively; WD Orbit is the range of projected physical separations at which a companion of that mass could be found around the white dwarf, measured in AU; MS Orbit is the range of projected physical separations at which a companion of that mass could be found around the main sequence progenitor, measured in AU.}} & \multicolumn{1}{c}{\bf Spectral} & \multicolumn{1}{c}{\bf $\mathbf{t_{\rm{tot}}}$} & \multicolumn{1}{c}{\bf $\mathbf{50\%}$~$\mathbf{J}$} & \multicolumn{2}{c}{\bf $\mathbf{50\%}$~$\mathbf{M}$} & \multicolumn{1}{c}{\bf $\mathbf{50\%}$~$\mathbf{T}$} & \multicolumn{2}{c}{\bf WD Orbit} & \multicolumn{2}{c}{\bf MS Orbit}\\
\multicolumn{1}{c}{\bf Dwarf} & \multicolumn{1}{c}{\bf Class} & \multicolumn{1}{c}{\bf [Gyr]} & \multicolumn{1}{c}{\bf [mag]} & \multicolumn{2}{c}{\bf [$\mathbf{M_{\rm{Jup}}}$]} & \multicolumn{1}{c}{\bf [K]} & \multicolumn{2}{c}{\bf [AU]} & \multicolumn{2}{c}{\bf [AU]}\\
\hline
WD0046$\,+\,$051 & DZ & 3.8 & 22.7 & $\,\,7$ & $^{0}_{1}$ & 290 & $\;\;$13 & 190 & $\;$3 & 48\\
WD0115$\,+\,$159 & DQ & 1.7 & 22.0 & $\,\,8$ & $^{1}_{1}$ & 380 & $\;\;$46 & 675 & $\;$11 & 160\\
WD0141$\,-\,$675 & DA & 3.1 & 22.2 & $\,\,8$ & $^{2}_{1}$ & 320 & $\;\;$29 & 483 & $\;$7 & 123\\
WD0148$\,+\,$467 & DA & 2.5 & 21.9 & $\,\,10$ & $^{2}_{1}$ & 390 & $\;\;$48 & 457 & $\;$14 & 138\\
WD0208$\,+\,$396 & DAZ & 2.6 & 22.5 & $\,\,9$ & $^{1}_{1}$ & 360 & $\;\;$50 & 758 & $\;$13 & 194\\
WD0341$\,+\,$182 & DQ & 3.3 & 22.9 & $\,\,10$ & $^{2}_{1}$ & 360 & $\;\;$57 & 801 & $\;$16 & 222\\
WD0435$\,-\,$088 & DQ & 4.1 & 22.7 & $\,\,9$ & $^{1}_{2}$ & 320 & $\;\;$28 & 408 & $\;$9 & 124\\
WD0644$\,+\,$375 & DA & 2.1 & 22.4 & $\,\,8$ & $^{1}_{1}$ & 360 & $\;\;$46 & 652 & $\;$17 & 236\\
WD0738$\,-\,$172 & DZ & 2.4 & 22.0 & $\,\,7$ & $^{1}_{1}$ & 320 & $\;\;$27 & 379 & $\;$7 & 96\\
WD0912$\,+\,$536 & DCP & 3.0 & 22.1 & $\,\,9$ & $^{1}_{2}$ & 350 & $\;\;$31 & 419 & $\;$7 & 93\\
WD1055$\,-\,$072 & DC & 3.3 & 22.6 & $\,\,9$ & $^{1}_{1}$ & 340 & $\;\;$36 & 503 & $\;$8 & 103\\
WD1121$\,+\,$216 & DA & 2.3 & 22.2 & $\,\,8$ & $^{2}_{1}$ & 350 & $\;\;$40 & 605 & $\;$9 & 134\\
WD1134$\,+\,$300 & DA & 0.37 & 21.9 & $\,\,3$ & $^{1}_{0}$ & 350 & $\;\;$46 & 664 & $\;$9 & 127\\
WD1236$\,-\,$495 & DA & 1.4 & 21.9 & $\,\,8$ & $^{0}_{2}$ & 400 & $\;\;$49 & 987 & $\;$9 & 185\\
WD1344$\,+\,$106 & DAZ & 2.5 & 22.0 & $\,\,13$ & $^{0}_{2}$ & 440 & $\;\;$60 & 865 & $\;$14 & 208\\
WD1609$\,+\,$135 & DA & 2.8 & 22.5 & $\,\,10$ & $^{2}_{1}$ & 380 & $\;\;$55 & 642 & $\;$10 & 117\\
WD1626$\,+\,$368 & DZ & 2.2 & 22.8 & $\,\,8$ & $^{1}_{1}$ & 360 & $\;\;$48 & 535 & $\;$13 & 141\\
WD1633$\,+\,$433 & DAZ & 3.0 & 22.3 & $\,\,10$ & $^{2}_{2}$ & 370 & $\;\;$45 & 533 & $\;$10 & 123\\
WD1647$\,+\,$591 & DAV & 0.91 & 22.0 & $\,\,5$ & $^{0}_{1}$ & 350 & $\;\;$33 & 372 & $\;$7 & 77\\
WD1900$\,+\,$705 & DAP & 1.1 & 22.2 & $\,\,5$ & $^{1}_{0}$ & 330 & $\;\;$39 & 452 & $\;$8 & 89\\
WD1953$\,-\,$011 & DAP & 2.1 & 21.7 & $\,\,8$ & $^{1}_{1}$ & 360 & $\;\;$34 & 509 & $\;$7 & 111\\
WD2007$\,-\,$219 & DA & 1.4 & 22.4 & $\,\,7$ & $^{1}_{1}$ & 370 & $\;\;$55 & 831 & $\;$12 & 189\\
WD2007$\,-\,$303 & DA & 1.7 & 22.3 & $\,\,7$ & $^{2}_{1}$ & 360 & $\;\;$46 & 834 & $\;$12 & 224\\
WD2047$\,+\,$372 & DA & 0.89 & 21.8 & $\,\,6$ & $^{1}_{0}$ & 390 & $\;\;$54 & 202 & $\;$12 & 46\\
WD2105$\,-\,$820 & DA & 1.3 & 21.1 & $\,\,9$ & $^{1}_{1}$ & 430 & $\;\;$51 & 639 & $\;$11 & 137\\
WD2140$\,+\,$207 & DQ & 4.4 & 21.6 & $\,\,13$ & $^{3}_{0}$ & 370 & $\;\;$38 & 542 & $\;$13 & 181\\
WD2246$\,+\,$223 & DA & 1.7 & 22.0 & $\,\,9$ & $^{1}_{1}$ & 400 & $\;\;$57 & 835 & $\;$11 & 157\\
WD2326$\,+\,$049 & DAZ & 1.1 & 21.8 & $\,\,6$ & $^{1}_{1}$ & 370 & $\;\;$41 & 396 & $\;$9 & 89\\
WD2359$\,-\,$434 & DA & 2.9 & 22.3 & $\,\,7$ & $^{1}_{1}$ & 310 & $\;\;$24 & 433 & $\;$4 & 82\\
\hline
\end{tabular}
\end{table}

\begin{table}
\caption{Recent imaging searches for wide companions\label{bddesert}}
\begin{tabular}{llccr@{ - }lc}
\hline
\multicolumn{1}{l}{\bf Survey\tablenote*{(1) \citet{mz2004}; (2) \citet{fbz2005}; (3) \citet{akm2007}; (4) \citet{ldm2007}; (5) \citet{ncb2008}; (6) \citet{hbc2009}}} & \multicolumn{1}{l}{\bf Targets}  & \multicolumn{1}{c}{\bf Number} & \multicolumn{1}{c}{\bf Limit} & \multicolumn{2}{c}{\bf Separation} & \multicolumn{1}{c}{\bf Frequency of}\\
& & \multicolumn{1}{c}{\bf of Targets} & \multicolumn{1}{c}{[\bf $\mathbf{M_{\rm{Jup}}}$]} & \multicolumn{2}{c}{\bf [AU]} & \multicolumn{1}{c}{\bf Companions}\\
\hline
(1) & G K M & 102 & $>$~12 & 75 & 300 & $1\%\pm1\%$\\
& & 178 & $>$~30 & 140 & 1200 & $0.7\%\pm0.7\%$\\
& & & 5-10 & 75 & 300 & $<3\%$\\
(2) & White Dwarfs & 261 & $>$~52 & 100 & 5000 & $<0.5\%$\\
& & 86 & $>$~21 & 50 & 1100 & $<0.5\%$\\
(3) & M7-L8 & 132 & $>$~52 & 40 & 1000 & $<2.3\%$\\
(4) & F G K M & 85 & 13-40 & 25 & 250 & $<5.6\%$\\
(5) & A F G K M & 60 & $>$~4 & 20 & 100 & $<20\%$\\
(6) & White Dwarfs & 29 & $>$10 & 60 & 200 & $<8\%$\\

\hline
\end{tabular}
\end{table}

\section{Summary}

From these results, tentative conclusions regarding the frequency of 
substellar and planetary mass companions to white dwarfs and their main 
sequence progenitors at wide separations can be made (we recognise that the 
DODO survey contains a relatively small number of targets). These conclusions 
assume that no common proper motion companions are confirmed around the 
remaining white dwarf requiring a third epoch image. Firstly, using the $90\%$ 
completeness limits, the DODO survey can detect companions with effective 
temperatures $>540$~K around {\em{all}} targets. Therefore, we suggest that 
$<4\%$ of white dwarfs have substellar companions with effective temperatures 
$>540$~K between projected physical separations of $60-200$~AU, although for 
many fields this applies to smaller ($\sim13$~AU for WD~$0046+051$; 
\citealt{bch2008}) and larger ($\sim1000$~AU) projected physical separations. 
This corresponds to projected physical separations around their main sequence 
progenitors ($1.5-8\,M_{\odot}$, i.e., spectral types F5--B5) of $20-45$~AU, 
although again for many fields these limits apply to smaller ($\sim3$~AU for 
WD~$0046+051$; \citealt{bch2008}) and larger ($\sim230$~AU) projected physical 
separations. For the same range of projected physical separations around both 
white dwarfs and main sequence progenitors and using the $50\%$ completeness 
limits, we suggest that $<7\%$ of white dwarfs and their main sequence 
progenitors have companions with masses above the deuterium burning limit 
($\sim13\,M_{\rm{Jup}}$), while $<8\%$ have companions with masses 
$>10\,M_{\rm{Jup}}$.

The results from the DODO survey can be compared to the results from other 
imaging surveys for wide substellar and planetary mass companions to white 
dwarfs and main sequence stars (Table~\ref{bddesert}). In particular, our 
results are consistent with those of \citet{mz2004} and \citet{ldm2007}. The 
DODO survey results can also be compared to complimentary recent MIR searches 
for {\em{unresolved}} substellar and planetary mass companions to white dwarfs 
(e.g., \citealt{mkr2007}). A recent MIR photometric survey of 27 white dwarfs 
using the Spitzer Space Telescope and IRAC, which was performed by 
\citet{fbz2008}, was sensitive to the entire known T dwarf sequence. Their 
observations place similar limits ($<4\%$) on the frequency of such companions 
to white dwarfs, but at smaller separations (with some overlap) compared to 
the DODO survey. 

\section{Future Work}

Since the DODO survey contains a relatively small number of targets, 
increasing the number of white dwarfs observed will increase the likelihood of 
directly detecting a planet around a white dwarf. Additional white dwarfs have 
recently been discovered in the local neighbourhood (e.g., \citealt{shb2007}, 
\citealt{hso2008}, \citealt{shb2008}, \citealt{sjh2009}), which has provided 
the opportunity to extend the DODO survey. First epoch images of 10 new white 
dwarfs within $\sim40$~pc have already been obtained with Gemini North and 
\textit{NIRI}, and a proposal to obtain the second epoch images for these 
white dwarfs will be submitted next year.

\begin{theacknowledgments}
Based on observations obtained at the Gemini Observatory, which is operated by 
the Association of Universities for Research in Astronomy, Inc., under a 
cooperative agreement with the NSF on behalf of the Gemini partnership: the 
National Science Foundation (United States), the Science and Technology 
Facilities Council (United Kingdom), the National Research Council (Canada), 
CONICYT (Chile), the Australian Research Council (Australia), Minist\'{e}rio 
da Ci\^{e}ncia e Tecnologia (Brazil) and Ministerio de Ciencia, Tecnolog\'{i}a 
e Innovaci\'{o}n Productiva (Argentina).
\end{theacknowledgments}

\def\mnras{MNRAS}
\def\aap{A\&A}
\def\apj{ApJ}
\def\apjl{ApJ}
\def\apjs{ApJS}
\def\pasp{PASP}
\def\aj{AJ}


\end{document}